\documentclass[preprint,amsmath,amssymb,superscriptaddress]{revtex4}
\usepackage{epsfig}
\usepackage{expdlist}
\usepackage{amsmath}
\usepackage{amssymb}

\begin{document}

\def\beqa{\begin{eqnarray}}
\def\eeqa{\end{eqnarray}}
\def\beqn{\begin{equation}}
\def\eeqn{\end{equation}}

\def\A{A}
\def\ac{\alpha}
\def\bc{\beta}
\def\gc{\gamma}
\def\dc{\delta}
\def\ab{{\bar \ac}}

\def\n#1{n_{#1}}
\def\nz#1{n^{(0)}_{#1}}
\def\N#1{N_{#1}}
\def\qc#1{{q_{#1}}}
\def\cn#1{{c_{#1}}}
\def\nm#1{{\bar n}_{#1}}
\def\ng{n}

\def\t{t}
\def\Dt{\Delta \t}
\def\tp{\tau}

\def\P{{\cal P}}
\def\R{R}
\def\Rb{{\dot\R}}
\def\Ra{A}
\def\Rd{D}
\def\Rf{F}
\def\S{S}
\def\F{F}
\def\T{T}

\def\M{M}
\def\lm#1{\lambda_{#1}}
\def\cl{r}

\def\u{u}
\def\v{v}
\def\w{w}
\def\z{z}
\def\us{u}
\def\ut{\tilde u}
\def\vs{v}
\def\vt{\tilde v}

\def\x{x}
\def\cp{s}

\title{The criticality of multiplicative processes}

\author{Bernard Gaveau}
\affiliation{UPMC-SU, 4 place Jussieu, 75005, Paris, France (Retired)}
\author{Marc-Thierry Jaekel}
\affiliation{Laboratoire de Physique Th\'eorique de l'ENS, CNRS, PSL, UPMC-SU, 24 Rue Lhomond, 75005 Paris, France}
\author{Jacques Maillard }
\affiliation{ACELIDES, 71 rue Bichat, 75010 Paris, France}
\author{Jorge Silva}
\affiliation{UPMC-SU, 4 place Jussieu, 75005, Paris, France (Retired)}


\date{\today}

\begin{abstract}
Keeping in view applications to numerical simulations of the evolution
of a nuclear reactor core around criticality,  
we use a general mathematical framework for describing the evolutions of multiplicative processes (processes involving particle creation) both in particle generations and in time. 
This framework allows us to obtain, within a same formalism, two corresponding estimates of the multiplication factor which describes the growth of particle numbers at large times.
We obtain the relative positions of both estimates with respect to each other and to criticality. 
These relations may show particularly useful
when simulating in a realistic way the monitoring of nuclear cores
in subcritical states, such as is the case for Accelerator Driven Systems (ADS).
More generally, this study applies to various multiplicative processes which can be found in nature.

\end{abstract}
\maketitle

\section{Introduction}

The development of new designs represents an important challenge for the future of nuclear energy production.
The complexity of the physics involved and the strong constraints imposed by practical realizations,
in particular security issues, imply that the approaches dedicated to the development of original designs
heavily rely on modelizations and simulations and that the latter must be made as faithful as possible. 
Simulations appear crucial for studying the feasibility
of innovative designs mixing different technologies. Such is the case in particular of  Accelerator Driven Systems (ADS) which associate a nuclear core, which burns a fissile material, with a proton source, producing by spallation the neutrons which initiate the nuclear reactions chains \cite{Gaveau1,KUCA,MYRRHA}. Characteristic features of these systems, which play an important role with respect to security issues, are their operation below criticality and the ability to monitor their power using the proton source \cite{KUCA1,KUCA2}.

\bigskip
In an approach which is commonly used to develop models and simulations of nuclear cores and to study their properties with respect to criticality, one focuses on the behavior of averaged neutron fluxes at large time scales \cite{Reuss}.
Such an approach is well suited to the study of the neutronic evolution of homogeneous nuclear cores in quasi stationary states,
such as those building the nuclear plants which are operated for power production.
This shows particularly useful when studying power production, fissile material regeneration or fission products accumulation on a long term. 
In that case, criticality refers to the multiplicative character of the averaged neutron flux over the whole core, and is represented by a
multiplication factor (equivalently, by a $\rm{k\_{eff}}$ coefficient).  The latter, which characterizes the global state of the nuclear core at large time scales, can be obtained in Monte Carlo simulations as the largest eigenvalue of a time independent matrix describing the effect of fissions on 
successive generations of neutrons within the nuclear core \cite{Reuss,Gaveau2}. 

\bigskip
In the case of ADS however, one must consider the correlated time evolutions of the different elements composing the core, paying attention to the spatial distributions of neutron fluxes and matter contents. Simulations with a time specific approach appear to be necessary \cite{Williams,Pazsit}, as the transient properties of neutron fluxes and the position of the spallation source relative to the nuclear core now play an important role.
This is exhibited in experimental ADS, where measurements of neutron fluxes show that 
the properties of the proton source used for producing spallation neutrons strongly influence
the time and spatial distributions of neutrons within the core \cite{KUCA3}.
In general, an approach with an explicit time dependence may be required, for instance when assessing the effect of security devices or when studying the impact of the protron source on the critical properties of the nuclear core \cite{KUCA4,KUCA5}.
In that case, the time evolution of neutron fluxes is better described by a time evolution operator acting on spatial densities.
The multiplication factor characterizing the growth of neutron fluxes is defined from this time evolution operator. The question then arises of relating the corresponding properties of neutron fluxes, which can be locally measured,
to the parameters which characterize the global state of the nuclear core in a time independent way, in particular with respect to criticality \cite{KUCA5}.

\bigskip
Due to intrinsic complexities, the generation and the time dependent approaches are usually not followed simultaneously within a same formalism.
This constitutes a drawback for modelizations and simulations of the time monitoring
 of nuclear cores, close to criticality and with significant spatial inhomogeneities as in ADS. 
 In particular, ambiguities arise when trying to relate kinetic parameters, which characterize the global state of the nuclear core and which can be obtained from Monte Carlo simulations, to experimental measurements performed on neutron fluxes at chosen locations \cite{KUCA3,KUCA5}. 
 These difficulties show to be critical when trying to assess the efficiency of security devices,
such as absorber bars, or when studying the consequences of a power breakdown in ADS. 
In such cases, a bridge between the different approaches would be much helpful, as it would allow one to compare
the kinetic parameters, like the multiplication factor, which are obtained from different types of simulations and to apply these results to experimental situations.

\bigskip
The fundamental processes occurring in a nuclear reactor may be characterized as multiplicative processes, i.e. processes involving particle creation, hence numbers of particles which vary in time.
These fundamental processes, although characterized by cross-sections which slowly vary in time, 
build the nuclear reaction chains which, together with the system geometry, ultimately determine 
the global parameters characterizing the system (produced power, neutron fluxes, matter contents, ...), which 
become time dependent. Simulations, either in deterministic or Monte Carlo approaches, rely on numerical computations
and hence require discretized representations of time and space, under the form of an elementary time interval, or time step, and elementary space cells. 
The errors induced by this discretization, both statistically and systematically, are controlled to a required level of precision
by diminishing the sizes of the elementary time step and space cells. The resulting increase in computation time 
is ultimately limited by available computing capacities. The important improvement in velocity and memory capacities reached by computers in the last decades has significantly increased the precision attainable by simulations. 
But present simulation codes still need to be strongly optimized in order to reach the sensitivity level allowing a comparison with experimental measurements. Simulating systems at very low subcriticality even requires too small time steps
to allow one to satisfactorily estimate the corresponding multiplication factor. 
The situation is even more critical for ADS, as optimization leads to adopting different time steps for describing the evolutions of different parts of the system
(proton source, nuclear core), thus entailing tradeoffs which may significantly affect the precision which can be reached.

\bigskip
When focusing on the large time behavior of a nuclear core, these difficulties can be circumvented in Monte Carlo simulations by using the generation approach, that is by generating neutrons
according to some characteristics and by following all particles (including neutrons and fission products), which are successively generated by interacting with the different materials building the nuclear core. Parameters like the multiplication factor are then deduced from a transfer matrix, representing the evolution of the system along successive generations of neutrons, without relying on an explicit time step dependence \cite{Gaveau2}.
Nevertheless, the nature of the simulation code still allows one to recover time dependencies, at the expense of degrading the precision level. The same approach can be applied to ADS and, near criticality, the different time steps may be optimized so that to allow one to obtain some time dependent paramaters with a satisfactory precision level. It should then become feasible to compare the estimations of the kinetic parameters which are given by the generation and the time dependent approaches.

\bigskip
The multiplication factor (or equivalently the $\rm{k\_{eff}}$ coefficient), which is used to characterize the global state of a system with respect to criticality,
plays a crucial role when addressing safety and efficiency issues.
As two different estimates can be given of the multiplication factor, that is, according to the generation approach or to 
the time dependent approach, it becomes important to determine their relative positions with respect to each other and with respect to criticality.
The aim of this article is to follow both  approaches simultaneously within a same formalism, and hence to prove that both estimates of the multiplication factor can be used equivalently near criticality
and, more precisely, to determine their relative positions in a rigorous way.

\bigskip
The formalism which is used here is the one underlying Monte Carlo simulations, i.e. the Master equation approach. 
This formalism appears in non-equilibrium statistical mechanics in order to describe the evolution of many body systems \cite{Keizer,Kubo,Kubo2,Gaveau}. 
It is very briefly presented in section 2, together with the processes which are specific to  nuclear reactors (absorption, diffusion, fission). The equation describing the time evolution of averaged numbers of particles is deduced.
In this approach, time is discretized, as well as the particles phase space. 
In section 3, generations of particles are defined: 
initially, all particles belong to generation 0  until the first creation event, after which they belong to generation 1, etc...
The generation evolution equation is given by a matrix with non negative elements, for which the Perron-Frobenius theorem applies \cite{Grantmacher}, insuring the existence of a highest positive eigenvalue controlling the spectrum of the matrix. 
Generation criticality is defined in terms of this eigenvalue in section 4. 
Both definitions of criticality (in terms of time evolution and of generation evolution) are then  proved to be equivalent.
The behavior around criticality is also studied in this section.

\section{Multiplicative processes}

In this section, we first present the basic definitions and the assumptions which allow one to apply the Master equation approach to
the time evolution of multiplicative processes, in the context of discretized time and phase space.
Specifying the different elementary processes affecting a neutron in a nuclear reactor core, one then applies the Master equation approach
to the description of neutron densities within the core and obtains the equations which describe the time evolution of the numbers of neutrons in the different elementary cells.

\subsection{Cells and configurations}

We consider a finite set $\A$, with elements denoted by $\ac, \bc, ... \in \A$. An element of $\A$ will be called a cell.
A configuration is the data of the occupation numbers $\{\n{\ac}\}$
of the cells $\ac$, where $\n{\ac}$ is a positive integer, the number of particles in cell $\ac$.
We consider a stochastic process $\{\N{\ac}(\t)\}$ associated with the evolution of occupation numbers $\{\n{\ac}\}$ in time $\t$ and define the corresponding probability
\beqa
\label{Probability}
\P(\{\n{\ac}\},\t) \equiv {\rm{Prob}}\{\N{\ac}(\t)=\n{\ac}, \forall \ac\in\A\}
\eeqa   
The time variable $\t$ is either discrete or continuous. In the former case, a time step $\Dt$ is chosen and fixed once for all.
To take into account the initial configuration $\{\nz{\ac}\}$, we define the conditional probability
 \beqa
\P(\{\n{\ac}\},\t| \{\nz{\ac}\}) \equiv {\rm{Prob}}\{\N{\ac}(\t)=\n{\ac}, \forall \ac\in\A | \N{\ac}(0) = \nz{\ac}, \forall \ac\in\A \}
\eeqa  
We shall use the obvious convention $\P(\{\n{\ac}\},\t)=0$ if one of the $\n{\ac}$ is strictly negative. 

\subsection{Elementary processes}

We define a multiplicative stochastic process $\{\N{\ac}\}$ in discrete time as a Markov process
with transition probabilities in one time step
\beqa
\label{transition_probabibility}
{\rm{Prob}}(\{\N{\ac}(\t+\Dt)\} = \{\n{\ac}^\prime\} | \{\N{\ac}(\t)\} =  \{\n{\ac}\}) \equiv \R(\{\n{\ac}^\prime\} | \{\n{\ac}\})
\eeqa
which satisfies the following conditions:

(i) we assume that the various particles (belonging to the same or different cells) do not interact with each other, 
{\it i. e.}  are independent. 
As a consequence, it is sufficient to specify the transition probabilities of the elementary processes
 in one time step for a given particle in a given cell $\bc$ (for every $\bc$)

(ii) we define $\R(\{\qc{\ac}\} | \bc)$  the probability that a particle in cell $\bc$ at time $\t$ creates,
 at time $\t+\Dt$, $\qc{\ac}$ particles in the cell $\ac$, 
for all $\ac$. $\{\qc{\ac}\}$ is a collection of  integers which are positive for $\ac\neq\bc$ and $\ge -1$ for $\ac=\bc$
 (destruction of one particle in $\bc$ and possible creation of particles in $\bc$).
 We assume that, if $\{\qc{\ac}\} \neq \{0\}$ {\it i.e.} at least one particle is created 
 in some cell or destroyed in $\bc$, the quantities $\R$ are proportional to $\Dt$ 
\beqa
\R(\{\qc{\ac}\} | \bc) \equiv \Rb(\{\qc{\ac}\} | \bc) \Dt, \quad {\rm{for}}\quad \{\qc{\ac}\} \neq \{0\}
\eeqa 
(iii) we assume that the probability that a particle in cell $\bc$ does not create any new particle and is not destroyed is a positive quantity, defined by
\beqa
\R(\{0\} | \bc) \equiv 1 - \sum_{\{\qc{\gc}\} \neq \{0\}}\R(\{\qc{\gc}\} | \bc) 
\eeqa
Hence, the transition probabilities in one time step $\R(\{\n{\ac}^\prime\} | \{\n{\ac}\})$ between a configuration $\{\n{\ac}\}$ at time $\t$ and a configuration $\{\n{\ac}^\prime\}$ at time $\t+\Dt$ for the stochastic process $\{\N{\ac}\}$ are specified by
\beqa
\label{1-transition_probability}
&&\R(\{\n{\ac}+\qc{\ac}\})| \{\n{\ac}\})
\equiv \sum_\bc \R(\{\qc{\ac}\} | \bc) \n{\bc}, \quad{\rm{if}}\quad \{\qc{\ac}\}\neq\{0\}\nonumber\\
&&\R(\{\n{\ac}\}| \{\n{\ac}\}) \equiv  1 - \sum_{\substack{\{\qc{\gc}\} \neq \{0\}\\\bc}}\R(\{\qc{\gc}\} | \bc) \n{\bc}
\eeqa

\subsection{Master equation and equations for the moments}

The evolution of the probability $\P(\{\n{\ac}\})$ (equation (\ref{Probability})) is given in discrete time by
the transition probability (\ref{transition_probabibility}) and equations (\ref{1-transition_probability}) 
\beqa
\label{master_equation}
\P(\{\n{\ac}\},\t+\Dt)&=& \left(1 - \sum_{\substack{\{\qc{\gc}\} \neq \{0\}\\\bc}}\R(\{\qc{\gc}\} | \bc) \n{\bc}\right)\P(\{\n{\ac}\},\t) \nonumber\\
&&+ \sum_{\substack{\{\qc{\gc}\} \neq \{0\}\\\bc}}  \R(\{\qc{\gc}\} | \bc) (\n{\bc}-\qc{\bc}) \P(\{\n{\bc}-\qc{\bc}\},\t)
\eeqa
(with $\P(\{\n{\ac}\}) = 0$ as soon as $\n{\ac}<0$ for one $\ac$).

For each cell $\ac\in\A$ the mean occupation number $\nm{\ac}(\t)$ and its correlations $\cn{\ac,\bc}(\t)$ are defined by
\beqa
&&\nm{\ac}(\t)\equiv<\N{\ac}(\t)> \equiv \sum_{\{\n{\gc}\}}\n{\ac}\P(\{\n{\gc}\})\nonumber\\
&&\nm{\ac,\bc}(\t)\equiv<\N{\ac}(\t)\N{\bc}(\t)> \equiv \sum_{\{\n{\gc}\}}\n{\ac}\n{\bc}\P(\{\n{\gc}\})\nonumber\\
&&\cn{\ac,\bc}(\t)\equiv \nm{\ac,\bc}(\t) -\nm{\ac}(\t) \nm{\bc}(\t)
\eeqa
The evolution equations for the mean occupation number and its correlations are deduced from the master equation (\ref{master_equation})
\beqa
\label{evolution_equations}
\nm{\ac}(\t+\Dt) &=&\nm{\ac}(\t) +  \sum_{\substack{\{\qc{\gc}\},\bc}}\R(\{\qc{\gc}\} | \bc) \qc{\ac}\nm{\bc}(\t)\nonumber\\
\nm{\ac,\bc}(\t+\Dt) &=& \nm{\ac,\bc}(\t) + \sum_{\substack{\{\qc{\gc}\},\dc}}\R(\{\qc{\gc}\} | \dc) \left(\qc{\ac}\qc{\bc}\nm{\dc}(\t) + \qc{\ac} \nm{\bc,\dc}(\t)+ \qc{\bc}\nm{\ac,\dc}(\t)\right)\nonumber\\
\cn{\ac,\bc}(\t+\Dt) &=& \cn{\ac,\bc}(\t) + \sum_{\substack{\{\qc{\gc}\},\dc}}\R(\{\qc{\gc}\} | \dc) \left(\qc{\ac}\qc{\bc}\nm{\dc}(\t) + \qc{\ac} \cn{\bc,\dc}(\t)+ \qc{\bc}\cn{\ac,\dc}(\t)\right)\nonumber\\
&&- \left(\sum_{\substack{\{\qc{\gc}\},\dc}}\R(\{\qc{\gc}\} | \dc)\qc{\ac}\nm{\dc}(\t)\right)\left(\sum_{\substack{\{\qc{\gc}\},\dc}}\R(\{\qc{\gc}\} | \dc)\qc{\bc}\nm{\dc}(\t)\right)
\eeqa
Let us remark that closed equations (\ref{evolution_equations}) have been obtained for the first and second moments of the random variable $\N{\ac}(\t)$. This is due to the fact that 
the particles do not interact with each other, so that the transition probability in the master equation  depends linearly on the number of particles (see (\ref{master_equation})).
  
\subsection{Specification of elementary processes}

From now on, the particle will be assumed to undergo one of the following elementary processes:

(i) Absorption

In this case, a particle in  cell $\bc$ is absorbed (and then disappears) in one time step,
 so that $\qc{\bc}=-1$ and $\qc{\ac}=0$ for $\ac\neq\bc$. 
The probability for absorption will be denoted by $\Ra_\bc$
\beqa
\R^{absorption}(\{\qc{\ac}\}|\bc) \equiv \Ra_\bc \prod_\ac \delta(\qc{\ac}+\delta_{\ac\bc})
\eeqa

(ii) Diffusion

In this case, a particle in cell $\bc$ is transmitted to a cell $\gc\neq \bc$, 
so that $\qc{\gc}=1$, $\qc{\bc}=-1$ and $\qc{\ac}=0$ for $\ac\neq\bc$ and $\ac\neq\gc$.
 The probability for diffusion will be denoted by $\Rd_{\gc\bc}$
\beqa
\R^{diffusion}(\{\qc{\ac}\}|\bc) \equiv \Rd_{\gc\bc} \prod_\ac \delta(\qc{\ac}+\delta_{\ac\bc}-\delta_{\ac\gc}), \qquad \Rd_{\bc\bc}\equiv0
\eeqa

(iii) Fission

In this case, a particle in cell $\bc$ creates $\qc{\ac} \ge 0$ particles in cell $\ac$ in one time step 
with $\sum_{\ac\in\A}\qc           {\ac}>0$, so that one has net creation.
Moreover, either $\qc{\bc}=-1$ (the particle in destroyed during the fission process) or 
 $\qc{\bc}\ge 0$ (the particle in cell $\bc$ is recreated during the fission process together with other particles).
The probability for fission will be denoted by $\Rf(\{\qc{\ac}\}|\bc)$ 
\beqa
\R^{fission}(\{\qc{\ac}\}|\bc) \equiv \Rf(\{\qc{\ac}\}|\bc), \qquad
 \Rf(\{0\}|\bc) \equiv 0
\eeqa

(iv) Nothing

The probability that a particle in cell $\bc$ remains in cell $\bc$ doing nothing in one time step is thus (the sum of the probabilities of processes (i), (ii), (iii) and (iv) must be $1$) 
\beqa
\label{diagonal_R}
&&\R^{nothing}(\{\qc{\ac}\}|\bc) \equiv \R(\{0\}|\bc) \prod_\ac \delta(\qc{\ac})\nonumber\\
&&\R(\{0\}|\bc)= 1 -\Ra_\bc - \sum_{\gc}\Rd_{\gc\bc} -\sum_{\{\qc{\gc}\}}\Rf(\{\qc{\gc}\}|\bc)
\eeqa
This last quantity is assumed to be positive and $\Ra$, $\Rd$ and  $\Rf$ to be  proportional to the time step $\Dt$, which should be a small quantity, so that (\ref{diagonal_R}) will be close to $1$ for small $\Dt$.
Finally, we denote by $\R$ the substochastic matrix defined by
\beqa
\label{matrix_R}
\R_{\ac\bc} &\equiv&\sum_{\{\qc{\gc}\}}\left(
\R^{absorption}(\{\qc{\gc}\} | \bc) + \R^{diffusion}(\{\qc{\gc}\} | \bc) + \R^{nothing}(\{\qc{\gc}\} | \bc)\right)\qc{\ac} \nonumber\\
&&+ \delta_{\ac\bc}(1 - \sum_{\{\qc{\gc}\}}\R^{fission}(\{\qc{\gc}\} | \bc))\nonumber\\
&=& \Rd_{\ac\bc} + \R(\{0\}|\bc)\delta_{\ac\bc}\nonumber\\
&&\sum_{\ac}\R_{\ac\bc} \le 1
\eeqa
Then, $\R_{\ac\bc}$ describes the probability for a particle in a cell $\bc$ to diffuse to another cell $\ac$ or to stay in the same cell $\bc$ without doing anything (absorption or fission).

The time evolution of the mean occupation number (\ref{evolution_equations}) 
may then be rewritten in terms of a matrix $\S$ defined by
\beqa
\label{matrix_S}
&&\nm{\ac}(\t+\Dt) = \sum_\bc \S_{\ac\bc}\nm{\bc}(\t)\nonumber\\
&&\S_{\ac\bc} \equiv \delta_{\ac\bc} + \sum_{\{\qc{\gc}\}}\R(\{\qc{\gc}\} | \bc) \qc{\ac} = \R_{\ac\bc} + \F_{\ac\bc}\nonumber\\
&&\F_{\ac\bc} \equiv \sum_{\{\qc{\gc}\}}(\qc{\ac}+\delta_{\ac\bc})\Rf(\{\qc{\gc}\}|\bc)
\eeqa
Then, $\F_{\ac\bc}$ describes the mean number of particles produced by fission in a cell $\ac$, in one time step, by a particle originating from a cell $\bc$.  
The second term in the fission matrix $\F$ takes into account the fact that a particle destroyed by fission in a cell $\bc$  recreates $\qc{\bc} + 1$ particles if they are produced in the same cell $\bc$ (the diagonal part of the matrix $\R$ accounts for the destruction of one particle in cell $\bc$). 

The time evolution of the second moment (\ref{evolution_equations}) may similarly be rewritten
in terms of the matrix $\S$
\beqa
\nm{\ac,\bc}(\t+\Dt) &=& \sum_{\gc,\dc}\left(\S_{\ac\gc}\delta_{\bc\dc} + \S_{\bc\gc}\delta_{\ac\dc}
-\delta_{\ac\gc}\delta_{\bc\dc}\right)\nm{\gc,\dc}(\t)\nonumber\\
&&+ \sum_{\{\qc{\gc}\},\dc}\R(\{\qc{\gc}\}|\dc)\qc{\ac}\qc{\bc}\nm{\dc}(\t)
\eeqa

Note that the time step $\Dt$ is constrained by the positivity of the probability $\R_{\ac\bc}$
\beqa
\label{constraint_Dt}
0 \le \R(\{0\}|\bc) \le 1 \quad \Rightarrow \quad 0 \le \Dt \le {1\over \displaystyle{\max_\bc\left({\dot \Ra}_\bc + \sum_\gc {\dot \Rd}_{\gc\bc} + \sum_{\{\qc{\gc}\}}{\dot \Rf}(\{\qc{\gc}\}|\bc)\right)}}
\eeqa 

Equations (\ref{matrix_S}) are the main result of this section. 
They explicitly give the time evolution of the spatial distribution of neutrons
once the cross sections of the elementary processes affecting neutrons in the reactor nuclear core are known.
The latter, together with the system geometry, i.e. the definition and matter contents of the elementary cells, completetely determine the matrix $\S$,
which summarizes the probabilities of the elementary processes occuring in each cell.

\section{Evolution of generations}

In this section, we define the notion of generation for multiplicative processes and the transfer matrix $\T$ describing the corresponding evolution equation. We also obtain the relation between the two matrices, respectively  $\S$ and $\T$, which describe the evolution of the system
with respect to time and generations respectively.

\subsection{Generations}
Suppose that at time $\t=0$ a configuration $\{\nz{\ac}\}$ is given.
 The particles of this configuration will be called particles of generation $0$. 
Let $\{\N{\ac}\}$ ($\t \equiv k\Dt$, $k$ integer) the stochastic process starting from $\{\nz{\ac}\}$.
We consider a given particle of generation $0$ at time $\t=0$ and follow it until it produces a fission event:
 the particles which are produced at this fission event are called particles of first generation.
 Let $\ng_1(\ac)$ the number of particles of first generation produced in cell $\ac$ by the particles of generation $0$. 
 Clearly, these particles are not produced at the same time.
 It can also happen that a particle of generation $0$ is absorbed before producing any fission event,
 so that it produces no particle of generation $1$. 

We define recursively $\ng_i(\ac)$ as the number of particles of generation $i$ produced (by fission) in a cell $\ac$.
 We consider the fission event (if any) produced by a particle of generation $i$
at some time. This fission event produces particles in various cells which are called particles of generation $i+1$. 
The total number of particles of generation $i+1$ in a cell $\ac$ will be denoted by $\ng_{i+1}(\ac)$. By definition also, $\ng_0(\ac) \equiv \nz{\ac}$. 

Let $\T_i(\ac|\bc)$ be the number of particles of generation $i$ produced in a cell $\ac$ by a single particle of generation $0$ in cell $\bc$. The matrix $\T$ with elements $\T_1(\ac|\bc)$ will be called the generation transfer matrix. It may be that $\T_i(\ac|\bc)$ is $0$, but it is also clear that
\beqa
\label{generation_evolution}
&&\ng_i(\ac) = \sum_\bc \T_i(\ac|\bc) \ng_0(\bc)\nonumber\\
&&\T_i(\ac|\bc) = (\T^i)_{\ac\bc}
\eeqa
One also has for one generation step
\beqa
\ng_{i+1}(\ac) = \sum_\bc \T_{\ac\bc}\ng_i(\bc)
\eeqa 
which is the evolution equation for generations.

\subsection{Transfer matrix for generations}

If a particle starts at time $\t=0$ in cell $\bc$, the probability that in $n$ time steps it propagates to a 
cell $\ac$ (or stays in $\bc$ if $\ac=\bc$), without being absorbed or producing fission, is given by $(\R^n)_{\ac\bc}$ 
where $\R$ is the matrix defined in equation (\ref{matrix_R}). 
Then, the probability that a particle propagates in any time steps from a cell $\bc$ to a cell $\ac$ 
without being absorbed or producing fission is given by $(\sum_{n=0}^\infty \R^n)_{\ac\bc} = ({1\over1-\R})_{\ac\bc}$. 
Then, the element $\T_{\ac\bc}$ of the generation transfer matrix, {\it i.e.} the mean number of particles produced 
in a cell $\ac$ by a particle originating from a cell $\bc$, is given by 
\beqa
\label{transfer_matrix}
 \T_{\ac\bc} = (\F{1\over1-\R})_{\ac\bc}
\eeqa  
Relation (\ref{transfer_matrix}) for the generation transfer matrix may also be obtained from the following reasoning.
 In one time step, a particle in a cell $\bc$ either diffuses to a cell $\ac$ without producing fission, 
with probability $\R_{\ac\bc}$,
or produces $\qc{\ac}$ particles ($\qc{\bc}+1$ if $\ac=\bc$) by fission with probability $\F(\{\qc{\ac}\}|\bc)$ (see (\ref{matrix_S})). 
Then, a particle of a given generation in a cell $\ac$ either comes from a particle of the same generation from another cell $\gc$ or  
has been produced by fission in $\ac$.
So that the number of particles $\T_{\ac\bc}$ produced by fission in cell $\ac$ by a particle from a cell $\bc$ must satisfy
\beqa
&&\T_{\ac\bc} = \sum_\gc \T_{\ac\gc}\R_{\gc\bc} + \F_{\ac\bc}\nonumber\\
\Leftrightarrow \quad &&\T = \F{1\over 1-\R}
\eeqa
 $1-\R$ must be invertible for relation (\ref{transfer_matrix}) to make sense. We briefly show that this is the case
 when the matrix $\R$ is irredicible, due its definition (\ref{matrix_R}).
From the definition (\ref{matrix_R}), one has $\sum_\ac\R_{\ac\bc} =1 - \Ra_\bc -\sum_{\{\qc{\gc}\}}\F(\{\qc{\gc}\}|\bc) \le 1$
 and assuming that there is at least one absorption or fission process, one deduces that there is at least one $\bc$ 
for which $\sum_\ac\R_{\ac\bc} <1$. 
Assuming that $\R$ is irreducible ($\forall \gc,\dc, \exists n: (\R^n)_{\gc\dc}\neq 0$),
 there follows that the strict inequality holds for every element $\gc\in\A$ for some 
$\R^n$ ($\forall \gc\in\A, \exists n:  \sum_\ac(\R^n)_{\ac\gc}\le\sum_\ac\R_{\ac\bc} <1$).
But then, $\R$ cannot have 1 as an eigenvalue.
 For in that case, the corresponding left eigenvector $\u$ would satisfy contradictory properties 
\beqa
&&\ab: \quad |\u_\ab|\equiv \max_\ac|\u_\ac|\nonumber\\
&&|\u_\ab| = |\sum_\ac\u_\ac(\R^n)_{\ac\ab}| \le\max_\ac|\u_\ac| (\sum_\ac(\R^n)_{\ac\ab})< \max_\ac|\u_\ac| 
\eeqa

Relation (\ref{transfer_matrix}) shows that the generation transfer matrix corresponds to a partial resummation 
of the time evolution,
$\R^n$ corresponding to the free evolution of a particle in $n$ time steps.
 The partial re-summation provides equations which extend the time evolution equations (\ref{matrix_S}) 
beyond the perturbative regime (in $\Dt$). Indeed, while the probabilities of elementary processes are
 of the order of the time step $\Dt$,
which is bounded according to equation (\ref{constraint_Dt}),
in contrast the generation transfer matrix is of order $1$
\beqa
&&\R \equiv 1 + {\dot \R}\Dt + O(\Dt^2) , \qquad \F\equiv {\dot \F}\Dt + O(\Dt^2)\nonumber\\
&&\T = \F\sum_{n=0}^\infty ({\dot \R} \Dt)^n = -{{\dot \F}\over {\dot \R}} + O(\Dt)
\eeqa
The following relation holds between the time evolution matrix $\S$
(\ref{matrix_S}) and the generation transfer matrix $\T$ (\ref{transfer_matrix})
\beqa
\label{relation_ST}
\S = \R + \F = 1 - (1-\T)(1-\R)
\eeqa 
Relation (\ref{relation_ST}) constitutes the main result of this section. It allows one to compare the evolutions of the
system, either in time, with the matrix $\S$, or in generations with the transfer matrix $\T$.
Let us note that, whatever the transfer matrix $\T$, the time evolution matrix $\S$ always remains close to 1
for small time step $\Dt$.

\section{Criticality}

In this section, we define the multiplication factors which can be associated with the evolution matrices $\S$ and $\T$,
and derive the relation connecting these two definitions, using relation (\ref{relation_ST}) between matrices $\S$ and $\T$. 
We also derive a more explicit expression for this relation close to criticality.

\subsection{Notations}
Let $\M$ be a matrix with positive elements. According to Perron-Frobenius theorem, 
there exists an eigenvalue $\lm{\M}>0$
with an eigenvector with positive components, such that all eigenvalues $\lm{r}$ of $\M$
 are complex numbers with $|\lm{r}| \le \lm{\M}$. 
If $\M$ is irreducible, in the sense that $\M^n$ has all elements strictly positive for $n$ large enough,
 $\lm{\M}$ is non degenerate and its eigenvector has all its components strictly positive.
Moreover, the eigenvalue $\lm{\M}$ is given by
\beqa
&&\lm{\M} = \max_{\v>0} \min_k{(\M\v)_k\over\v_k}, \qquad (\v>0 \quad \Leftrightarrow \quad \v_k >0,\forall k) 
\eeqa
For a stochastic matrix $\M$ ({\it i.e.} $\sum_k\M_{kl} =1, \forall l$), then $\lm{\M}=1$.

Eigenvalues of the matrix $\M$ will be arranged by decreasing order of moduli
 $\lm{\M} \equiv \lm{O} > |\lm{1}| \ge ...\ge |\lm{n}|$ and left and right eigenvectors with eigenvalue $\lm{r}$
 will be denoted by $\u^r(\M)$ and $\v^r(\M)$ respectively and normalized according to $\sum_k \u^r_k\v^r_k=1$,
 so that powers of $\M$ will read
$\M^n_{kl} = \sum_{r}\lm{r}^n\u^r_k\v^r_l$.

\subsection{Criteria for criticality}

Two a priori different notions of criticality may be given for a multiplicative process.
 The first one depends on the time evolution of the mean occupation number (\ref{matrix_S}) 
while the second one depends on the evolution of generations (\ref{generation_evolution}). 
The large time asymptotics for the solution of the time evolution equation for the mean occupation number 
and for the generation number respectively give
\beqa
\label{asymptotics}
\nm{\ac}(N\Dt) &\simeq& \lm{\S}^N \u^0_\ac(\S)(\sum_\gc \v^0_\gc(\S)\nz{\gc}), \qquad {\rm{for}} \quad  N >> 1\nonumber\\
\ng_I(\ac) &\simeq& \lm{\T}^I \u^0_\ac(\T)(\sum_\gc \v^0_\gc(\T)\nz{\gc}), \qquad {\rm{for}} \quad I >> 1
\eeqa
Recall that criticality is defined as the condition for stationarity of particle numbers at large time, {\it i.e.} $\lm{\S} =1$ or $\lm{\T}=1$. Hence, $\lm{\S} <1$ (resp. $\lm{\S} >1$) or $\lm{\T}<1$ (resp. $\lm{\T} >1$) correspond to subcriticality (resp. supercriticality). 
Two different criteria for criticality are thus obtained, according  to  the largest eigenvalue $\lm{\S}$ or $\lm{\T}$ which is chosen for discussing the asymptotic behavior of particle numbers.

One must note a fundamental difference between the two criteria. 
The first criterion relies on a comparison between two time scales,  a small one $\Dt$ (see (\ref{constraint_Dt})),
 necessary to define the infinitesimal probabilities (proportional to $\Dt$) of  elementary processes, 
and a large one $\t \equiv N\Dt$, which describes the time of evolution of the system.
 The ratio of these two time scales is assumed to be a very large number $N$, so that the evolution in time 
can in fact be described by differential equations, leading to  exponential behaviors with respect to time 
($\sim exp( \kappa\t/\Dt)$).      
The second criterion does not depend on any time scale and involves numbers only, as the generation number $I$. 
In fact, the latter does  not need to be very large.
This property is important for practical purposes, as the second criterion describes the evolution
 of the system with respect to its energy content and is more pertinent for determining the criticality of the system,
for example in nuclear reactors.   

 In the second criterion, the number $N$ corresponds to an upper bound in the partial resummation over all time steps. The very large but finite value of $N \equiv {\t/\Dt}$ gives an approximation to the infinite sum. The latter is controled by the ratio between the overall evolution time $\t$ of the system  and the infinitesimal time step $\Dt$. The finite $N$ approximation then does not play a role in establishing the asymptotic regime for the second criterion but may  affect the eigenvalue which determines the critical behavior of generations.  
The precise relation between the two criteria is studied  in next sections.

\subsection{Comparison of criticality criteria}

First, the following property is easily proven:

(i) if $\lm{\S}<1$, then $\lm{\T}<1$

(ii) if $\lm{\T}\ge 1$, then  $\lm{\S}\ge 1$

This property is a consequence of the following relations implied by the definitions of the mean particle and generation numbers and by their asymptotic behaviors (\ref{asymptotics}) 
\beqa
&&\sum_{I,\ac} \ng_I(\ac) \le \sum_{N,\ac} \nm{\ac}(N\Dt)\nonumber\\
&&\sum_{I,\ac} \ng_I(\ac) < \infty \quad \Leftrightarrow \quad \lm{\T} < 1\nonumber\\
&&\sum_{N,\ac} \nm{\ac}(N\Dt) < \infty \quad \Leftrightarrow \quad \lm{\S} < 1
\eeqa

In fact, one can prove a more refined  result:

(i)  $\lm{\T}=1$ if and only if  $\lm{\S}=1$

(ii)  $\lm{\T}<1$ if and only if  $\lm{\S}<1$, in which case $\lm{\T}\le\lm{\S}<1$

(iii) $\lm{\T}>1$ if and only if  $\lm{\S}>1$, in which case $1<\lm{\S}\le\lm{\T}$

Property (i) is a direct consequence of relation (\ref{relation_ST}) between matrices $\S$ and $\T$, recalling that $1-\R$ is invertible. We now prove properties (ii) and (iii).

We define $\w\equiv (1-\R)^{-1}\v^0(\T)$, so that
\beqa
&&\F \w = \T\v^0(\T) = \lm{\T}(1-\R)\w\nonumber\\
&&\S \w = \R\w + \F\w = \lm{\T}\w + (1-\lm{\T})\R\w
\eeqa
Noting that $1-\R$, $\v^0(\T)$, hence $\w$, and $\R$, hence $\R\w$, have only positive elements, one first deduces that
\beqa
\label{inequality1}
\lm{\T} \le 1 \quad &&\Rightarrow \quad (\S\w)_\ac \ge \lm{\T}\w_\ac\nonumber\\
&&\Rightarrow \quad \lm{\S} = \max_{\v>0} \min_\ac{(\S\v)_\ac\over\v_\ac}\ge \min_\ac{(\S\w)_\ac\over\w_\ac}\ge \lm{\T}
\eeqa
On the other hand, defining $\z \equiv (1-\R)\v^0(\S)$, one has
\beqa
\T\z &=& \F\v^0(\S) = \lm{\S}(1-\R)^{-1}\z - \R (1-\R)^{-1}\z\nonumber\\
&=&(\lm{\S}-1)(1-\R)^{-1}\z  + \z
\eeqa
Then, noting that $\v^0$, $\z$ and $\R^n$ have only positive elements, so that 
$ ((1-\R)^{-1}\z)_\ac \ge \z_\ac$, one also deduces that
\beqa
\label{inequality2}
\lm{\S} \ge 1 \quad &&\Rightarrow \quad {(\T\z)_\ac\over\z_\ac} \ge \lm{\S}\nonumber\\
&&\Rightarrow \quad\lm{\T}=\max_{\v>0} \min_\ac{(\T\v)_\ac\over\v_\ac} \ge \min_\ac{(\T\z)_\ac\over\z_\ac} \ge  \lm{\S}
\eeqa
Equations (\ref{inequality1}) and (\ref{inequality2}) may be summarized as
\beqa
\label{inequality3}
\lm{\T} < 1 \quad &&\Rightarrow \quad  \lm{\T}\le \lm{\S} < 1\nonumber\\
\lm{\S} > 1 \quad &&\Rightarrow \quad \lm{\T} \ge\lm{\S} > 1
\eeqa

Then, one knows that for any matrix $\M$, $(1-\x\M)^{-1} \equiv \sum_{n=0}^\infty\x^n\M^n$ is a convergent series provided that $|\x|<1/\max({\rm{Spec}}(\M))$. If $\M$ is a matrix with positive elements, the radius of convergence is given by $1/\lm{\M}$.
Now, assuming $\lm{\S}\le 1$ and noting that $\R$ and $\T$ have only positive elements, one has
\beqa
\label{inequality4}
(1-\x\S)^{-1}  &=& (1-\R)^{-1}\left(1 - (\x-1)\R(1-\R)^{-1} - \x \F(1-\R)^{-1}\right)^{-1}\nonumber\\
\lm{\S} < 1 \quad &&\Rightarrow \quad \sum_{n=0}^\infty\left((\x-1)\R(1-\R)^{-1} + \x\T\right)^n < \infty, \quad \forall \x: 1 \le |\x| < {1\over\lm{\S}}\nonumber\\
&&\Rightarrow \quad \sum_{n=0}^\infty\left(\x\T\right)^n < \infty, \quad \forall \x:  0 \le |\x| < {1\over\lm{\S}}\nonumber\\
&&\Rightarrow \quad {1\over\lm{\S}} \le {1\over\lm{\T}}
\eeqa
Inequalities  (\ref{inequality3}) and (\ref{inequality4})
provide parts (ii) and (iii) of the result.

\subsection{Perturbation around criticality}

We consider now a system which depends on an additional parameter $\cp$, 
so that it is described by $\cp$-dependent matrices ($\R(\cp), \F(\cp)$) and ($\S(\cp)$, $\T(\cp)$), 
and which remains near criticality ($\lm{\S}(0)=\lm{\T}(0) =1$ for $\cp=0$).
We show that, in the neighborhood of criticality, a simple relation exists between the eigenvalues 
of the two matrices $\S(\cp)$ and $\T(\cp)$  
\beqa
\label{criticality_relation}
\lm{\T}(\cp) -1  = \cl  (\lm{\S}(\cp) -1), \qquad \cl > 1
\eeqa
As a consequence of the theorem of previous section, the proportionality coefficient  $\cl$ must be greater than $1$.
We shall determine its value and show that it only depends on the matrix $\R$ and on the eigenvectors 
of $\S$ and $\T$ evaluated at criticality.

To simplify the notations we shall write $\u^0(\S(\cp)) \equiv \us(\cp), \u^0(\T(\cp)) \equiv \ut(\cp)$ 
and $\v^0(\S(\cp)) \equiv \vs(\cp), \v^0(\T(\cp)) \equiv \vt(\cp)$
\beqa
\label{cp_eigenvectors}
&&\us(\cp) \S(\cp) = \lm{\S}(\cp)\us(\cp), \qquad \S(\cp) \vs(\cp) = \lm{\S}(\cp)\vs(\cp)\nonumber\\
&&\ut(\cp) \T(\cp) = \lm{\T}(\cp)\ut(\cp), \qquad \T(\cp) \vt(\cp) = \lm{\T}(\cp)\vt(\cp)
\eeqa
 One also has from equation (\ref{relation_ST})
\beqa
\label{cp_relation_ST}
&&(1-\S(\cp)) = (1-\T(\cp))(1-\R(\cp))\nonumber\\
\Rightarrow \quad &&\us(0)=\ut(0)\nonumber\\
&&\vs(0)= (1-\R(0))^{-1}\vt(0)
\eeqa

Assuming differentiability in the parameter $\cp$, we obtain from equations (\ref{cp_eigenvectors})
 (eigenvectors are normalized)
\beqa
\label{cp_derivative1}
{d\lm{\S}(\cp)\over d\cp}|_{\cp=0} = \left(\us(0){d\S\over d\cp}|_{\cp=0}\vs(0)\right){1\over\us(0)\vs(0)}\nonumber\\
{d\lm{\T}(\cp)\over d\cp}|_{\cp=0} = \left(\ut(0){d\T\over d\cp}|_{\cp=0}\vt(0)\right){1\over\ut(0)\vt(0)}
\eeqa
One deduces from (\ref{cp_relation_ST})
\beqa
\label{cp_derivative2}
&&{d\T\over d\cp} =\left({d\S\over d\cp} + (\T-1){d\R\over d\cp}\right)(1-\R)^{-1}\nonumber\\
\Rightarrow \quad &&\ut(0){d\T\over d\cp}|_{\cp=0}\vt(0) = \us(0){d\S\over d\cp}|_{\cp=0}\vs(0)
\eeqa
From equations (\ref{cp_derivative1}) and (\ref{cp_derivative2}), one deduces the following relation 
between the variations of the two eigenvalues determining criticality
\beqa
{d\lm{\T}(\cp)\over d\cp}|_{\cp=0} ={d\lm{\S}(\cp)\over d\cp}|_{\cp=0}
 {\us(0)\vs(0)\over\us(0)(1-\R(0))\vs(0)}
\eeqa
Relation (\ref{criticality_relation}) is thus deduced in the vicinity of criticality, with the following value
 for the proportionality coefficient
\beqa
\label{ratio}
\cl &=& {\us(0)\vs(0)\over\us(0)(1-\R(0))\vs(0)} = \sum_n \left({\us(0)\R(0)^n\vs(0)\over\us(0)\vs(0)}\right)^n\nonumber\\
\us(0): &&\us(0) = \us(0)\S(0)=\us(0)\T(0)\nonumber\\
\vs(0): &&\vs(0)=\S(0)\vs(0)=(1-\R(0))^{-1}\T(0)(1-\R(0))\vs(0)
\eeqa 
The  proportionality coefficient $\cl$ is in effect greater than $1$.
 Moreover, the general properties of the two eigenvalues characterizing criticality,
 which have been briefly stated in a previous section, may now be discussed explicitly. 
One sees from equation (\ref{ratio}) that the ratio $\cl$ of their distances to $1$ is indeed proportional
 to $\tp/\Dt$, where $\tp \equiv 1/{\dot \R}(0)$ is a characteristic time for evolution, without fission,
 of the system at criticality. The ratio $\cl$ may be seen as the mean number of time steps occurring
 during an elementary propagation process (without  fission), or else as the inverse of the probability
 of propagation without fission. 
Equations (\ref{criticality_relation}) and  (\ref{ratio}) show that when this probability decreases, 
then the distance  to criticality increases. Also, the second criterion appears to be more efficient than
 the first one when determining the distance to criticality.

\section{Conclusion}

In this article, using a suitable formalism, we have described the evolution of a nuclear reactor core both in terms 
of neutron generations and in time. These two descriptions have led to two corresponding estimates of the multiplication factor, hence to two different criteria for criticality.
These two estimates have nonetheless been shown to be tightly related, one being bounded by the other near criticality.
This result confirms the intuitive picture, that both definitions of criticality should be equivalent, namely that the number 
of neutrons should increase exponentially in time if and only if it does so in terms of generations.
More precisely, we have obtained a rigorous comparison of the multiplication factors defined as the largest eigenvalues of the
matrices describing the evolution of neutron fluxes, both with criticality and with respect to each other.

\bigskip
This result provides the justification for extending the application of the generation approach to Monte Carlo simulations to ADS and for using
the deduced global parameters characterizing the system for a comparison with criticality.  
Amplification power and efficency of the whole system can thus be directly and simply connected with a $\rm{k\_{eff}}$ coefficient which
can be computed using the generation approach.
This gives confidence in the ability to use global criteria for maintaining a secure evolution of subcritical systems \cite{Xenofontos,Xenofontos2} and strengthens the argument in favor of the safeness capabilities of ADS designs.

\bigskip
Moreover, the formalism used here allows one to simultaneously follow the evolution of a nuclear core in the generation and the time dependent approaches. This allows one to connect the paramaters which characterize the global state of a nuclear core in a time independent way to the time dependent properties of neutron fluxes. This should help one to fill the gap usually separating the values for global parameters which are obtained from simulations and those which are deduced from direct measurements, in the case of heterogeneous systems such as ADS \cite{KUCA3,KUCA5}. 
This should help to significantly improve the comparison between modelizations and experimental realizations of ADS, and hence
to simulate and test realistic design in a way which remains close to practical setups and to technical requirements.

  \bigskip
  Let us finally remark that applications of the present study are not limited to the case of the multiplicative processes determining the evolution of nuclear reactor cores. 
Multiplicative processes also often appear in various chemical or biological reaction chains, for instance in molecular biology, genetics or population evolutions. Due to their multiplicative character, the processes underlying these systems lead to
equations describing their time evolution which are similar to those governing neutron fluxes in a nuclear reactor core. 
The formalism developed here could 
show helpful for their modelization and lead to new hindsights for their understanding.

\section{Acknowledgements}

The authors are much grateful to N. Catsaros, P. Savva,  M. Varvayanni, T. Xenofontos for collaborations on the application of simulation codes to nuclear reactors and to C.H. Pyeon, M. Yamanaka for 
stimulating and fruitful discussions on the experimental realization of ADS.



\end{document}